**Analyzing digital politics: Challenges and experiments in a dual perspective**

Géraldine Castel[1], Genoveva Vargas-Solar[2], Javier Espinosa-Oviedo[3]

*1. Univ. Grenoble Alpes, ILCEA4,*
*38000 Grenoble, France*
*geraldine.castel@univ-grenoble-alpes.fr*

*2. Univ. Grenoble Alpes, CNRS, Grenoble INP*, ILCEA4,*
*38000 Grenoble, France*
*genoveva.vargas@univ-grenoble-alpes.fr*

*3. Delft University of Technology*
*Julianalaan 134, 2628 BL Delft, Netherlands*
*javier.espinosa@tudelft.nl*

*RÉSUMÉ. Au cours de ces dix dernières années, les réseaux sociaux sont devenus un élément central de la vie politique. Cependant, pour ceux qui s'intéressent à l'analyse des stratégies de communication des partis et des candidats en période électorale, les conséquences de l'introduction d'Internet dans la sphère politique ont été mitigées. En effet, si la recherche, la consultation et l'archivage de documents originaux relatifs à une campagne spécifique sont devenus plus faciles, plus rapides et réalisables à plus grande échelle, ouvrant ainsi un eldorado prometteur pour la recherche dans ce domaine, l'étude des campagnes en ligne a aussi inévitablement introduit de nouveaux défis techniques, méthodologiques et juridiques qui se sont révélés de plus en plus complexes à résoudre pour les chercheurs en sciences humaines et sociales.*

*Le présent document propose donc de fournir un retour d'expérience ainsi qu'une validation expérimentale d'un projet pluridisciplinaire consacré à l'analyse comparative des campagnes politiques sur les réseaux sociaux à l'approche des élections au Parlement européen de 2014 en France et au Royaume-Uni. En plus d'observations formulées du point de vue des humanités sur les problématiques liées à un tel projet, le présent document présente des résultats expérimentaux concernant trois des phases du cycle de vie de la collecte des données : la collecte, le nettoyage et le stockage. Il en résulte une base de données prête à être analysées selon différents angles afin d'aider à traiter le sujet abordé en sciences politiques.*

*ABSTRACT. Social networks have become in the last decade central to political life. However, to those interested in analysing the communication strategies of parties and candidates at election time, the introduction of the Internet into the political sphere has proved a mixed blessing. Indeed, while retrieving, consulting, and archiving original documents pertaining to a specific campaign have become easier, faster, and achievable on a larger scale, thus opening up a promising El Dorado for research in this area, studying online campaigns has also inevitably introduced new technical,*

*methodological and legal challenges which have turned out to be increasingly complex for academics in the humanities and social sciences to solve on their own.*

*This paper therefore proposes to provide feedback on experience and experimental validation from a multidisciplinary project called POLIWEB devoted to the comparative analysis of political campaigns on social media in the run up to the 2014 elections to the European Parliament in France and in the United Kingdom. Together with observations from a humanities' perspective on issues related to such a project, this paper also presents experimental results concerning three of the data collection life cycle phases: collection, cleaning, and storage. The outcome is a data collection ready to be analysed for various purposes meant to address the political science topic under consideration.*

**1. Introduction**

To researchers interested in analysing the communication strategies of parties and candidates at election time, the introduction of the Internet into the political sphere has proved a mixed blessing. Indeed, while retrieving, consulting, and archiving original documents pertaining to a specific campaign such as posters, manifestos, broadcasts, speeches etc. have become easier, faster, and achievable on a larger scale, thus opening up a promising El Dorado for research in this area, studying online campaigns has also inevitably introduced new technical, methodological and legal challenges which have turned out to be increasingly complex for academics in the humanities and social sciences to solve on their own.

The use of information and communication technologies (ICT) in the political sphere is nowadays a key aspect for running electoral campaigns. In our work we focus on studying candidate practices in the context of electoral campaigns, and more precisely here, those for the elections to the European Parliament in 2014 in the UK and in France.

Trying to carry out the analysis of contents necessary to start forming conclusions on campaigning practices in both countries revealed a need for collaboration between several researchers for sharing out tasks and bringing together expertise from various disciplines. From the point of view of British civilisation and political sciences, an interdisciplinary approach was therefore a necessary scientific and technical prerequisite for being able to study the topic under consideration from data collected from heterogeneous sources. This challenge provided a contextualised framework guided by genuine application needs in which to develop solutions beyond the theoretical stage to Computer Sciences and Applied Mathematics.

This paper thus combines observations on collaborative projects like the POLIWEB one described here from a digital humanities' perspective to results concerning three of the data collections life cycle: collection, cleaning and storing. The final outcome is a data collection ready to be analysed for different purposes. In particular, in our experimental validation it has been used for comparing political campaign behaviour in France and the UK during the European elections in 2014. Accordingly, the remainder of the paper is organized as follows. Section 2 describes related work in the different fields involved in the project and more specifically on the use of ICTs in the political sphere and regarding data collection in computer science. Section 3 discusses the motivation and initial ideas for developing a digital solution to address a humanities' problem. Section 4 describes the various challenges the team had to address while section 5 offers suggestions for addressing the issues mentioned, including the presentation of a collection/storage solution for providing continuous data processing and curating data collections so as to make analytics processes possible. It also shows preliminary analytics results and conclusions driven from the collected data analytics. Finally, Section 6 concludes the paper and discusses future work.

## 2. Related work

From a political science perspective, the use of ICTs for political purposes is a relatively new research field as the first publications date from the end of the 1980s following experiments from a few isolated candidates in the United States, and then in a variety of other countries (Auty & Nicholas, 1998). Descriptive analyses of tools and practices were completed by reflections on the impact of such an evolution on the democratic processes of those countries (Grossman & Krueger, 1995; Margolis & Resnick, 2000).

Several campaigns were the object of scrutiny so as to reach a more detailed understanding of the topic (Foot & Schneider, 2008; Gibson & Ward, 2000; Vaccari, 2008). The continuous evolution of tools has been paralleled by a shift in attention from sites to forums (Marcoccia, 2006; Wojcik, 2003), blogs (Gadras, 2012; Greffet, 2007), and social networks (Jackson & Lilleker, 2011; Margaretten & Gaber, 2014).

However, in 2014, when the POLIWEB project was launched, most publications on the use of ICTs for political purposes in the European context focused on their appropriation, or lack of, by members of the European parliament as they were serving their mandates (Braghiroli, 2010; Stephen Coleman & Nathanson, 2005; Dai, 2007; Vesnic-Alujevic, 2013). Few had attempted to analyse, like Lusoli and Ward (Gibson & Ward, 2000) for the 2004 elections or Darren Lilleker and Al. (Jackson & Lilleker, 2011) had for the 2009 ones when blogs and sites were still prominent, the impact of the introduction of new technologies for campaigning for seats in the European parliament; and even less had been published on local initiatives as opposed to those launched by party headquarters.

In the field of computing, the following lines compare our work with approaches concerning data collections' design and preparation to support analytics processes. In the database domain, the problem of integrating databases from different sources is not new (Dong & Srivastava, 2013). Heterogeneous data integration on relational systems where heterogeneity was related to both data structure and semantics (Lara, Lausen, Arroyo, Bruijn, & Fensel, 2003) led to important results addressing schema integration, query rewriting and optimisation (Halevy, 2001). In most of these proposals, data providers (heterogeneous or not) are known in advance or discovered (Dong, Berti-Equille, Hu, & Srivastava, 2010) and integration is done assuming knowledge about the data structure (Cuevas-Vicenttín, Zechinelli-Martini, & Vargas-Solar, 2006), content, semantics (Osborne & Motta, 2015) and constraints.

Then, the emergence of new kinds of data providers as services introduced new challenges and particularly, a matching problem (Cuevas-Vicenttín et al., 2006). The assumption was that a query represented a data integration requirement which could be fulfilled by one or several data services, not known in advance, which should be looked up in registries (Meshkova, Riihijärvi, Petrova, & Mähönen, 2008). These approaches assumed, for example, the fact that the description of the data and content provided by services was stored as meta data, and the fact that services exported data in a pivot data model (Rekatsinas, Dong, Getoor, & Srivastava, 2015).

Data started to acquire "new" properties (more volume, velocity, variety) and with them emerged the need for building huge curated data collections out of data produced by different devices, under different conditions for later analysis (Adiba, Castrejón, Espinosa-Oviedo, Vargas-Solar, & Zechinelli-Martini, 2015; Labrinidis & Jagadish, 2012). The challenge was to collect data continuously (Ma, Wang, & Chu, 2013) and to ensure that collections could be used to perform analysis (Vargas-Solar, Espinosa-Oviedo, & Zechinelli-Martini, 2016): statistical, data mining, machine learning, deep learning and so on (Shah & Sheth, 1999). Works and tools include collection, cleaning, profiling and distributed storage (Barnaghi, Sheth, & Henson, 2013). Languages like, JaQL (Beyer et al., 2011), Pig Latin (Olston, Reed, Srivastava, Kumar, & Tomkins, 2008), as well as data cleansing and data mining techniques have been applied for this purpose. The objective was to complete data, to detect errors, ensure freshness and also to have views of its content (e.g., data types and value distribution, and possible correlations and dependencies among attributes (Park & Brenza, 2015). Resulting data collections were then stored according to different "sharding" techniques and sometimes they were correlated with other collections (Grolinger, Higashino, Tiwari, & Capretz, 2013). Data scientists could then decide which analytics techniques (Cugola & Margara, 2012) could be applied to extract information, infer models and knowledge from data collections. The challenge was to perform a continuous process as new data is harvested and as new insights are obtained about analysed data. In general, data processing is computationally expensive, and it requires storage and memory resources, since algorithms are greedy and require data in memory. So, works have emerged trying to study how to deploy solutions in architectures and environments which provide such resources. A great deal of research and technology has been devoted to parallel programming models, languages and environments deployed in architectures like the cloud, the grid or high-performance computing centres (Di Stefano, 2005).

Our work in POLIWEB thus addresses the construction of data collections giving a comprehensive view of their content, for supporting the decision making of data analysts and scientists willing to apply the most appropriate techniques which can lead to generate information and then knowledge. We maintain (without complete materialization) these views and all the sequence of data transformations done for preparing raw data to maintain some provenance properties without necessarily generating more data than those strictly required to ensure experiment reproducibility.

## 3. Motivation and initial ideas

The goal of the POLIWEB project from a political science perspective was to build on the work carried out by Roginsky (2012, 2014) for social media use by MEPs to try and find out how these practices translated at election time. As most political science analyses had concentrated on top-down communication strategies initiated from party headquarters to local campaigning teams or on voters' reactions to party initiatives, our work chose instead to concentrate on local practices at constituency level from individual candidates in the context of the 2014 elections to the European parliament both in France and in the UK. Moreover, as practices in France and the UK for digital campaigning to the European Parliament had been largely unexplored, it became a key interest for the POLIWEB project. Besides, the concern with social media for the 2014 election rather than the focus on sites, blogs or forums which had been privileged before was another breakaway from previous work. Finally, despite the coding grids developed in the past by HSS experts (Bastien & Greffet, 2009; Gibson & Ward, 2000; Goodchild, Oppenheim, & Cleeve, 2007; Kluver, Jankowski, Foot, & Schneider, 2007) to organise the collection of data and the reproduction of procedures, the processes involved were generally manual and therefore extremely time-consuming, which explains the reduced number of comparative studies between different countries (S. Coleman, 2006; Hoff, 2004; Lilleker, Koc-Michalska, Schweitzer, & Jacunski, 2011).

Significantly, the first researchers to work on the relationship between politics and technology, be they in the US John D.H Downing (Downing, 1989) or Jeffrey B. Abramson (Everts, Abramson, Arterton, & Orren, 1989) held PhDs in political science, as do such influential authors as Rachel Gibson (Fisher et al., 2017) or (A Chadwick, 2006; Andrew Chadwick & Howard, 2009). However, more and more projects in this area rely on the expertise of multidisciplinary teams. The PHEME consortium which aims to build tools made to assess the veracity of online claims is an example of such a trend as it gathers specialists of qualitative social media analysis, linguistics and semantics, as well as digital journalists, but also computer scientists focusing on large-scale Web data collection, storage and indexing, linked open data, graph based methods or large data analytics. Such collaborations are in keeping with most digital humanities' projects. Indeed, Dacos and Mounier in a report commissioned by the French Institute define digital humanities as an *"interdisciplinary dialogue on the digital dimension of research in the humanities and social sciences"* (Dacos & Mounier, 2015).

In the case of the POLIWEB project, this partnership emerged from the growing tension in research between on the one hand, the more and more widespread adoption of online communication by political actors, and on the other, the vulnerability of the data thus created. When the trend started at the end of the 20th century, the campaigning material shared online by candidates or parties was mostly replicas of offline documents. But progressively, they came to be replaced by endogenous contents with no offline equivalent such as digital debates on Youtube for example. Significantly, the blog post entitled '*7 things Hillary Clinton has in common with your abuela*' released in December 2015 by the candidate's team to try and appeal to Hispanic voters which triggered massive backlash on social media is now nowhere to be found. Moreover, if some candidates and parties are now careful to use for their online sites or blogs URL's with no reference to a specific election so as to be able to keep using them from one to the next, others do not, like the one launched for the 2014 campaign by the British Liberal party entitled http://www.europe4prosperity.org.uk/ or yet the http://www.choisirnotreeurope.com site of the French Socialist party which now both point to an error message when loaded.

The capacity to collect and store such data is thus no longer simply a matter of convenience but has become crucial from a scientific perspective as few traces if any remain after an election in traditional, offline archives, a phenomenon labelled 'information decay' by Mathew Ingram. This phenomenon is not limited to digital contents and campaigning material is by nature ephemeral but the instantaneous availability of such a variety of contents online sometimes conceals their potentially transient character.

Consequently, the POLIWEB project aimed at, beyond the topical issues mentioned above, looking into solutions for collecting online campaigning data so as to limit the loss of material over time and enable postfacto analysis. The ambition was to build a database containing quantitative and qualitative data as well as metadata collected from candidates' sites, Facebook and Twitter accounts, to organise data in a searchable database and to design a tool for basic statistical analysis and graph visualization to help extract meaning from the data available.

For this purpose, it sought to automate steps such as for instance the identification or the counting of certain elements so as to make the processes more comprehensive, more reliable, capable of dealing with larger amounts of heterogeneous data and to replicate them. A collaboration between data and humanities scientists within POLIWEB therefore made it possible to consider the design and implementation of semi-automatic collection, storage and analysis tools to manage data provided by heterogeneous Big Data. The production and release of Big Data collections is an important and growing challenge in society and research. To face this challenge, scientists from different disciplines have started to define new methodologies and experiments to open ways to exploit data in the best and most relevant way possible.

Data collections are provided in raw conditions, that is, with few descriptions of their content and with no validation of their quality (provenance, providers reputation, freshness, completeness, unicity). Data science communities have proposed cleansing processes, annotation techniques, models, aggregation and publication strategies to the scientific community. In Europe, the NESSI, for example, promotes the European Big Data Analytics Service Network (EBDAS) for facilitating the exchange of data among scientists and other consumers for preparing the next generation of products and applications targeting data analysis.

Current works underline the need for interaction between exact sciences and HSS to propose complementary solutions considering key aspects of the design, annotation, cleansing and exploitation of data considering the particular requirements of humanities. It was straightforward to associate in POLIWEB expertise on data for identifying significant correlations among data and develop models and understanding of election processes. Beyond the obvious challenge of proposing pluridisciplinary methodologies and solutions, concrete challenges included (i) the comparative binational, multi-support perspective desired in British civilization; (ii) the need for a database which could be queried according to different criteria which could include the number of parameters to consider for understanding trends.

POLIWEB represented the opportunity to reduce the distance between data producers and consumers through technical solutions guided by the requirements of humanities' methodologies and topical objectives. From a legal point of view, POLIWEB introduced the challenge of reasoning about legal aspects of data including their governance from the moment they are collected, until the moment they are used for understanding phenomena. From a broader perspective, it was also an opportunity to reflect upon epistemological issues raised by such projects.

## 4. Challenges addressed by the POLIWEB project

*4.1 Big Data in Digital Humanities*

Among the difficulties encountered by the team working on the POLIWEB project, a significant number were related to the fact that the corpus for the intended analysis qualified as Big Data.

The following definition of Big Data offered by the Cambridge dictionary though rather basic is nonetheless revealing in so far as it refers to "Very large sets of data [...] that can only be stored, understood, and used with the help of special tools and methods."[1] Such a reflection on methods and tools is indeed made capital by what Burt L. Monroe called the 'Five Vs of Big Data Political Science', i.e volume, velocity, and variety to which he adds vinculation and validity (Monroe, 2013).

Initially, from the 1990s, online campaigning was limited to a small number of candidates using their websites as mostly static shopping windows of their political offer. At the time, collecting and archiving data required little technical expertise for researchers. But the situation has evolved. According to Politico, from the first US presidential campaign announcement on March 23, 2015 through November 1, 2016, "128 million people on Facebook across the US generated 8.8 billion likes, posts, comments and shares related to the election." (Levy, 2016). Elections in France and the UK do not generate such levels of online activity, even less so those to the European parliament which generally fail to trigger much enthusiasm. However, even with the boundaries set for the POLIWEB project (only one test constituency analysed in each country, with candidates belonging exclusively to the five parties which had gained seats at the 2009 election), the team had to deal with the production of about a hundred and fifty candidates which amounted to thousands of publications, making it impossible for an individual or even a small group of researchers to manually collect, store and retrieve such quantities of data, all the more so as such data is constantly growing and of a heterogeneous nature. To quote Monroe:

*"Variety is one of the most challenging for social scientists to grapple with, as much Big Data comes in forms and data structures that do not map easily onto the rectangular spreadsheets of familiar statistics packages. Tweets are a useful example. They are delivered in structured text formats, either XML or JSON, that impose a nested tree structure on each tweet. The levels of this tree depend on whether the tweeter references a previous tweet. The number of fields varies depending on what the user provides and can include time stamps, geo-references, text, and images."*

In the context of POLIWEB, text, images, videos and metadata were collected.

The 'vinculation' mentioned by Monroe refers to yet another parameter, i.e the fundamentally interdependent nature of social data. For instance, if one decides to archive the contents of the public Facebook page of a candidate, what does it mean in practice? Collecting the data published by the candidate himself on his page exclusively? The whole contents of the page including what was posted by others such as comments? What about what he posts on pages which are not his? The links he posts to sites outside Facebook? And if one follows the trail, where to stop?

---

[1] *Big data* definition, Cambridge dictionary, Online version, October 2018.

The last V stands for validity and is related to the four others. With data so voluminous, so heterogeneous, so rapidly evolving and so complex in its architecture, how does one manage to extract reliable meaning?

*4.2 Black box effects*

This concern for validity is all the more prevalent for researchers in the humanities and social sciences involved in digital humanities' projects as they frequently necessitate the devolving of tasks previously managed manually at the individual level to colleagues or machines, which makes it more complex to ensure the validity of each step over the course of the whole process. For POLIWEB, detailed information on the data which was to be collected was handed over to the computer scientists who were in charge of the collection, storing and organising of the data and the end result, a workable database was provided accordingly. Regular dialogue between all participants took place but without a systematic sharing of most of the intermediary stages, which led to a black box effect in some areas.

For instance, once the database was designed, a test showed that a query for the posts from a British candidate who had been interviewed in person and declared to be very active on Facebook returned null results, suggesting that she had in fact not posted at all. Further examination revealed that, as the crawler used for the collection of the data had been blocked because of some privacy settings on her account, no data was retrieved for this candidate. While this was perfectly logical from a technical and legal perspective, it raised certain issues from a scientific one from the point of view of the contents specialist for whom the distinction between a non-publishing candidate and one whose data was not available was very important and could threaten the reliability of the results mentioned earlier. The problem was solved by differentiating the non-posting from the non-availability of data in the interface of the database. Yet is highlights the potential for black box phenomena in such initiatives.

These are to a certain degree inevitable as machines and specialists from other disciplines are precisely brought in to compensate for skills or tasks which were hitherto lacking. Yet the project demonstrated that being aware of such risks was necessary from the very beginning of a project to try and limit their impact when possible. For a researcher in the humanities and social sciences, this was not straightforward as not only was it crucial to take into account a variety of technical details, but this had to be done at a very early stage too, introducing such concepts as workflows and data management plans.

The difficulty of it came from a minimal grasp on computing issues, but also from a necessity to alter the order of the steps usually followed in the research process. In those disciplines, researchers are aware of an existing state of the art on the topic under consideration but often start working on a set of documents and let hypotheses emerge from close reading of the documents in an inductive approach. This proved much more arduous when dealing with Big Data. Indeed, unless some direction was defined before starting off working with the data, there was a real danger of losing one's way in its volume and complexity. But if it became imperative to formulate research questions clear enough to guide the collection and analysis, there was also a danger for the initial hypotheses to bias the end result. In this respect, the experience of the POLIWEB team echoed with the report written in 2015 by Gareth Millward, a historian involved in the Big UK Domain Data for the Arts and Humanities' project of the British Library who tried to investigate the information available to disabled people on the early world wide web thanks to the digital collections of the library:

*"The project encountered a number of issues with the database when trying to answer the initial research questions. […] Simply put, there was far too much information to answer the research questions I had originally formulated. It became obvious that the techniques used by traditional documentary historians rely on a lack of available evidence. That is, we tend to identify a question and source base, go to the archive, and then mine what we can until that vein is exhausted. This is possible because we have a relatively small amount of evidence which has survived. With the Archive, however, it is virtually impossible to create a corpus that is both small enough to be human-readable and provides a useful, relevant and representative sample across time."* (Millward, 2015).

Moreover, pre-defining a detailed workflow is all the more complex as investigation in the digital humanities often involves a research process which does not follow the usual linear model, due in great part to what ethnographers (Jouet & Le Caroff, 2013) call the "intellectual and technical DIY of online observation" following in the footsteps of (Kozinets, 2009). Indeed, even though digital humanities' projects have been carried out for about two decades, they still retain an experimental and empirical nature which requires a cyclical approach with to-and-fro movements and repeated phases of data collection, manipulation, building of hypotheses, preliminary results and technical adjustments.

But black box effects also raise the issue of the skills necessary for scholars in the humanities and social sciences to engage is this type of work. In 1968, French historian Emmanuel Le Roy Ladurie stated: "Tomorrow's historian will be a programmer or will be no more" (Le Roy Ladurie, 1968). Since then, the debate has been fierce between those, beyond the field of history, who defend the opinion that any self-respecting digital humanist should have some knowledge of Python, XML, R, or SQL and others who have worked very hard to achieve expertise in their own field and are unwilling to devote time, energy and money which they do not necessarily have to become what they see as second rate computer scientists.

*4.3 Methodological concerns*

However, an awareness of the technical considerations inherent to our POLIWEB project, if crucial, also needs to be completed by a reflection on the way technology interacts with the traditional methods of the disciplines involved. This relates to the new challenges raised in particular by Big Data but also to the delegation of tasks to machines or colleagues which had been in the past performed through processes which were largely personal, informal, intuitive and have to be made more explicit, precise and visible. Thus, it is necessary for digital humanities scientists to take a step back to look at one's habits through a different prism which is both stimulating and challenging.

In that respect, it seems useful to emphasize that experimental sciences and humanities, if they share a variety of characteristics, nonetheless emerge from different traditions and perspectives. Indeed, the terminology in French can be quite unflattering to the humanities, opposed as they are to 'Hard science' (*Sciences dures*) or 'exact science' (*Sciences exactes*) which seems to imply that the humanities would be 'soft' or 'approximate' science. In this context, it could be tempting to see digital humanities as an opportunity to demonstrate the opposite, to be dazzled by the amount of data available, the sophistication of the tools and to believe that in this new world, the practices of the past have become obsolete. Conclusions from the POLIWEB project are more ambivalent in this area. If indeed, the novelty of many aspects of digital research lead researchers to develop new approaches, it also offers an opportunity to rediscover the value of basic aspects of long-established methodologies in humanities.

For instance, researchers in the humanities are aware of the subjective nature of the documents they work with, be they speeches, poems, and know that a corpus is an artifact. Yet, the notion moves to the sidelines when dealing with *data*. The Latin etymology of the word *data* (Given) is itself misleading as explained by Christof Schöch in his article entitled "Big? Smart? Clean? Messy? Data in the Humanities" (Schoch, 2013). At the collection stage, the term *data* seems to refer to a simple compilation of words, figures, images available to be recorded and observed. At manipulation stage, *data* is understood as the objective outcome of a 'neutral' process. This led researchers such as Johanna Drucker to reject "assumptions of knowledge as observer-independent and certain, rather than observer co-dependent and interpretative. […] The concept of *data* as a given has to be rethought through a humanistic lens and characterized as *capta*, taken and constructed (Drucker, 2011). *Data*, whether collected or generated through manipulations is the result of a series of very human decisions and choices defining the final outcome of a data science process.

As mentioned earlier, if the decisions made at various points when data is collected and built into a corpus can affect the final outcome, so do the initial hypotheses. For Millward: "*Finding exactly what we expect to find ought to set off alarm bells for historians. At the same time, with no relevance searching at all, we run the risk of being unable to make any sense of the mass of data we have archive*d." (Millward, 2015). The balancing act can prove arduous to achieve. If knowing "the whole field" seems just as challenging in the era of Big Data as it could be for sociologists Glaser and Strauss in 1967, one can endeavour as they suggest to "develop a theory that accounts for much of the relevant behaviours" (Glaser & Strauss, 1967).

Yet defining relevance, reliability or representativeness is not straightforward either, especially as online data is so voluminous and diverse that any theory can be proved right depending on where in the data the researcher chooses to focus his or her lens, a phenomenon akin to what computer scientist and mathematician David Leinweber warns against when referring to " bunnies in the clouds": *"Everyone knew that if you did enough poring, you were bound to find that bunny sooner or later, but it was no more real than the one that blows over the horizon.*" (Leinweber, 2007).

This metaphor illustrates the statisticians' motto, sometimes unknown to digital humanists, that correlation is not causation and the risk of mistaking statistical and factual coincidences for meaningful trends. While the suspicious nature of the correlation between the divorce rate in Maine and per capita consumption of margarine highlighted on the Spurious Correlations' Website[2] seems easy enough to spot, examples from the research field can be far more confounding. When one is trying to find out if the use of ICTs by political candidates influences the outcome of an election and finds a correlation between the candidates with the most skills in using social networks for campaigning and those who win elections, is it because one is responsible for the other, a coincidence, or a correlation due to another factor, for example that the most active online are also generally younger and therefore potentially more attractive to some voters or that they have a more efficient communication strategy overall whatever the media?

This brings back to another truism of research in the humanities, that data whatever its nature, whatever the processes it went through, is not generated in a void and cannot be dissociated from a wider context. For instance, when trying to assess if French parties use Twitter more than British ones, one needs to be aware that there were twenty candidates on French lists and only ten on British ones, so that focusing on global averages might be misleading while individual averages might be more significant. If going back to the spreadsheet image used earlier, calculating an average for a mark and getting 25 is a problem in France where the highest mark is 20. Thus, a result can be mathematically correct, yet factually wrong because of various decisions made in the preceding stages. If the last decades demonstrated that for researchers in electoral strategies, ignoring the online dimension of political communication was no longer an option, the opposite is also true: to understand online data, a precise knowledge of offline factors is just as imperative.

### *4.4 Legal and Ethical Challenges*

The last series of challenges encountered by the POLIWEB team was legal, or ethical in some cases. Whereas in the past for instance, for a researcher to put in a folder at home a leaflet handed by a candidate on a market square was deemed acceptable, its online equivalent seems today far less so as databases are now regulated by two main regimes: copyright laws regarding their ownership and exploitation, and personal data protection laws, applicable when databases contain information on an identified or identifiable subject, which was the case here with political candidates. That a candidate's personal data such as the name of the school attended by his children should be protected goes without saying. But what about what he willingly chooses to share publicly on his blog, site or social network pages for clearly electoral purposes? Can it still be qualified as 'personal'? The law seems to make little distinction between both types of data whereas the difference is non-negligible.

---

[2] Spurious Correlations at http://tylervigen.com/spurious-correlations

Besides, such concerns were made all the more complex by the format of elections to the European parliament. As only the candidates at the top of their party's list stand a chance of being elected and those lists contain up to twenty persons, a significant number of them did not campaign actively in 2014 and were much more likely to use their private accounts on social networks to do so rather than setting up a distinct page. Hence, a candidate like Marlène Mourier, the UMP mayor of a French city, used the same Facebook page to broadcast information to her constituents, about her party, as a candidate to the European election of 2014, to support colleagues to the 2015 departmental elections but also to advertise for her favourite restaurant, with users sending wishes for her birthday. In such circumstances, it becomes very difficult to draw the line between the personal and the political as emphasised by Yves Gingras (GINGRAS, 2011) among others.

Moreover, the processing of personal data rests on the principle of 'prior consent' but this raises several issues. If one obtains the consent of a candidate over what he released on his Facebook page, should a researcher also get the consent of every single person who posted on it? What of the many, candidates included, who share data which is not theirs? What about documents such as the hundreds of pictures of rallies for instance whose authors are impossible to trace?

In France, sensitive data can be harvested if it has been made public by the data subject, but the definition of 'made public' in the case of online release is not consensual as explained by Michael Zimmer in his article entitled "But the data is already public: on the ethics of research in Facebook" (2010). While Cardon (2012) draws attention to "the paradox of private conversations held in public on social networks".

In the UK, personal data may be exempt from some of the data protection principles as long as the processing does not cause damage to the individual and as a result of the research itself does not identify any data subject (1998 Data Protection Act, Section 33). And indeed, anonymising a dataset can be a solution in some projects. Yet the technical difficulty of completely anonymising a dataset made up of online data when algorithms can identify users via their online traces has been demonstrated by Yves-Alexandre de Monjoye and his team (2015). Moreover, anonymisation is not an option when the goal of the project is to try and understand the communication strategies of specific candidates. And the attempt by the Conservatives in 2013 (Ballard, 2013) to erase from their own website but also from the Internet Archive past speeches and press releases suggests that parties and candidates may be reluctant to allow such archiving. Does the 'right to be forgotten' apply to public figures and to the data deliberately made public in that capacity?

Navigating the meanders of copyright laws can prove just as challenging. For example, who owns the data published on a candidate's timeline on Facebook? The candidate himself? Facebook? Both, as Facebook users grant the company a non-exclusive cession of copyrights? And what content exactly is protected by copyright? Whether a tweet for instance is copyright-protected is very much debated. Debate between the various stakeholders is still ongoing but the tension between the determination to engage into ethically responsible research while preserving access to resources for scientific purposes is tangible.

## 5. Addressing issues: Building and curating election campaign data collections and avenues for reflection

To address the Big issues raised in section 4, POLIWEB proposes data collection strategies specialized according to the type of data provider and implemented by services. As shown in figure 1, data collection is done according to different modes (push, pull) and at different rates particularly when data is produced continuously. Some providers are Web pages and blogs which contain and upload information, so we crawled the content using Web scrapping and crawling techniques and tools.

We designed and built a database from the online production of a selection of candidates based on their blogs, sites, Twitter and Facebook accounts to confront theories derived from an analysis of raw online data with field-work data. Hence, feedback from candidates was collected from online questionnaires and semi-guided interviews applying novel crowdsourcing techniques combined with privacy constraints. According to the type of data providers we used (i.e., Twitter, Facebook and official candidates web sites), we developed two general data collection strategies: on demand which use data pulling techniques, and continuous which supposed the production of data streams which were recurrently produced at some rate.

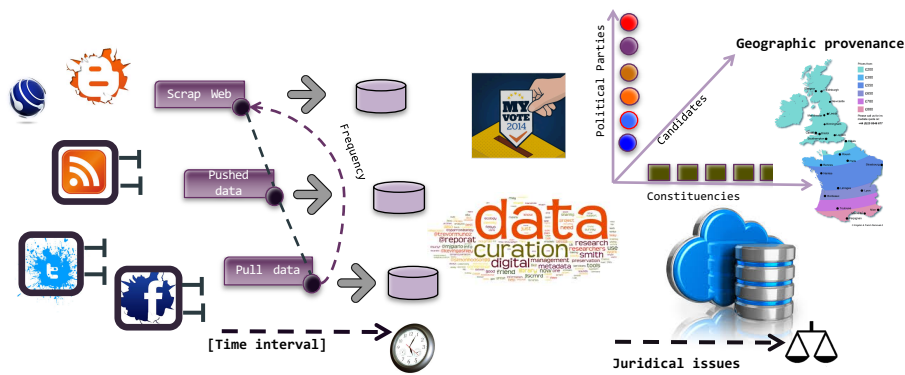

*Figure 1. Data collection and curation overview*

### 5.1 Collecting and archiving data

For collecting data, we assume that data providers are services implemented by a REST architecture or a SOAP API. These are public services and they can have specific SLA constraints, like the number of requests per hour (e.g., Twitter) or some authentication ones (e.g., Facebook). Other providers like pages or sites do not have explicit constraints but we assume that they are governed by privacy and authorship rights determined by their country of origin.

***On demand data providers*** have to be queried through a specific interface or crawled in order to harvest data. The frequency in which data is collected is specified in the program interacting with the provider. The consumer invokes the batch method through the network with its input data. The speed of the network introduces the transfer time cost which is determined by the data size and the network's conditions (i.e., latency and throughput). Depending on the type of network, it can have a monetary price (e.g., 3G) also determined by data size. Once the method invocation arrives to the hosting device, the service provider receives the request and associates a predefined method invocation price. Afterwards, the method instance processes the request during an execution time (i.e., method response time) which is determined by the method throughput given by the number of processed requests in a period of time (e.g., each minute) and the state of the device such as memory or CPU usage. The request implies the usage of the network interface, service provider and method execution. Those processes spend the battery of the device (i.e., battery consumption) entailing a battery cost. Finally, the output (i.e., method response) is sent back to consumer through the same network and, as input data aforesaid, output data contributes to data transfer time and to monetary cost. Both input and output data define the data size measure.

***Stream providers*** work under a subscription strategy. A continuous data provider exports an method subscribe() used by a consumer to start receiving streams at some rate and for a given period of time (e.g., by executing an unsubscribe() method, for a predefined period of time, until something happens). The general process is implemented to interact with this type of providers. The consumer invokes the continuous method through the network with its input data. Then, the method instance starts processing results and it sends the results every period of time (the so-called production rate). The production rate can be determined by consumer needs. For instance, "give my current position every five minutes" where 'five minutes' is the expected production rate. Produced data is then sent to the consumer who processes it immediately or after a threshold defined by the number of tuples received, or the elapsed time, or a buffer capacity. This threshold is named processing rate. Both production rate and processing rate impact the execution time cost, execution price cost, and battery consumption cost.

We assumed that providers are autonomous in the sense that they can modify their interfaces, authentication protocols and privacy and authorship rules whenever they want, and our data collection services must deal with these changes. We do not have information about the production rate of the streams and changes in Web pages and sites. In a first approach we tuned the collection manually, but we also collected information about services' behaviour to automatize the tuning process and ensure the collection of fresh non-redundant data.

We collected 30 Gigabytes of data about the European elections about candidates in the UK and in France. Data concerns campaigns of 12 parties and 100 candidates, and they concern only online activities reported on Twitter, Facebook and official sites, pages and blogs. We used JSON as data model and we then implemented document processing tasks to characterize the content of collected data.

*5.2. Using views for curating data collections*

Collected data composes raw data collections which must be analysed to get an abstract overview of the content in order to decide which cleaning and analytics techniques apply best to exploit them. The idea is not to transform data but to generate an abstract aggregated view and then eventually tag it with information which can be used for further data processing tasks which might generate transformed versions of this raw data. The view could be seen as a kind of schema in the relational world, but extracted a posteriori after having created a database.

In order to simplify, we assume that data is represented as tuples and documents under a JSON-like structure. Therefore, we define a view as a document that provides a description of every family of attributes of a raw document collection. For example, consider a collection of tweets from the European political campaign of candidates of the Labour party in the UK. A simplified version of these tweets has the following structure: <"user", "date", "time", "location", "content">, where almost all attributes are of type String and "content" can be of type String, Image, Video, Sound. Not all attributes are mandatory in every tweet and they can change of type from one tweet to another within the same collection.

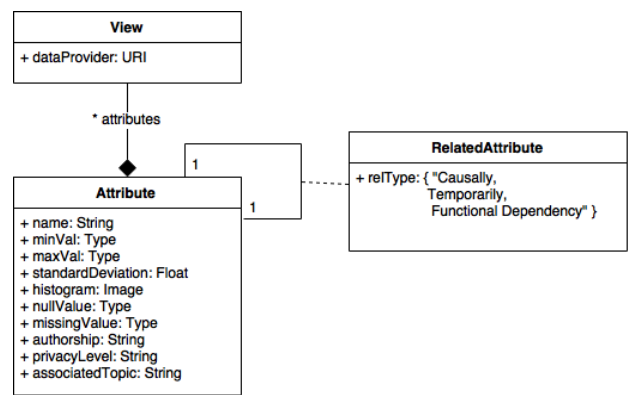

*Figure 2. UML class diagram of the View Curation Model*

As shown in the UML class diagram in figure 2, a View characterizes the content (document) provided by a given dataProvider as a set of attributes. For example, the view of our tweets collection consists of five classes of type Attribute, each class characterizing each of the attributes of a tweet, namely <"user", "date", "time", "location", "content">.

The concept Attribute provides a snapshot of a given attribute's values domain for a given dataset. For example, the attribute time in our political data set ranged between the official initial date of campaigns for European elections to the date of the official announcement of results. An attribute within the dataset has maximum and minimum values, a standard deviation of the values assigned to the attribute in the different documents collected in the dataset, and the variation of values across the dataset elements represented by a histogram. Within a dataset an attribute can have null and missing values which must be inferred in order to characterize its domain type as precisely as possible. Indeed, many data collections represent missing values by dummy values and therefore we want to represent those cases. For example, in our political tweets' collection, the attribute location was not always associated with a value. Since such values are inferred out by analysing a dataset, such values have associated precision probabilities which can measure uncertainty. In our political tweets, in some cases it was Twitter which associated an empty value.

A value of an attribute in a document in a dataset can have an author and it can be protected by an authorship license, some privacy level and it can belong to a thematic classification. In the case of the tweets and particularly photographs they belong to Twitter and to the user. For images this was a huge assumption because some candidates use photographs which do not have recognized authors. We avoided this problem, but this is part of open issues regarding juridical aspects studied in our work. Anyway, authorship and privacy level represent the condition in which the value of an attribute is produced and can be consumed.

Concerning thematic classification, in the case of our political tweets we used hashtags within the contents to classify the tweets by topic. The content of a tweet was parsed looking for hashtags which could help to group the tweet and then try to compare topics with key words of the political proposals or the official speech of the corresponding party. We could also see that some trendy topics on Twitter were not that trendy in sites, and blogs. Again, tweets' classification is associated to uncertainty and was associated to probabilistic measures.

Finally, an attribute can be related to other attributes within a document with different relationship types: functional dependency, temporal and causal dependencies. For example, in the case of a tweet, the temporal attribute of the tweet has a temporal dependency with the temporal attribute of the replies or retweets. Replies and retweets should always be published later as the initial tweet otherwise there is an error. Relationships are computed using numerical measures estimated from values. Some are more or less easy to identify. For example, we use hashtags to determine "semantic" similarity which we interpret as tweets using the same hashtags. Network science techniques provide a variety of strategies which can be used for providing a comprehensive view of the political campaigns done on Twitter and other social networks. Relationships which are deduced are tagged with probabilities and with measures which represent the influence of tweets towards others.

We associate a visual representation of a view which can be presented to a data scientist who can validate its contents. Once a data collection has been profiled with a view, the view can be stored, completed and modified and used for supporting a decision-making process of a data scientist who will decide which analytics techniques to apply in order to extract knowledge and drive conclusions and models about a given subject.

*5.3. Generating campaigns profiles for European candidates*

Based on our approach we built a system to analyse and compare campaigns in UK and France of the European elections in 2014. As shown in figure 3 recent advances in the field of social network analysis and data visualization could be put to use to chart the nature and direction of information flows from the original producer. From a research perspective, the solutions offered by the computer scientists proved extremely valuable for the analysis of the campaigns under consideration as it transformed a considerable volume of heterogeneous data into humanly manageable information. Access to synthetic data became immediate and customizable according to the specific angle being studied, be it individuals, parties, countries etc. and also greatly facilitating comparative work. As for the tools presented in figure 4, they enabled the identification of trends which would have been lost in the mass of data otherwise, such as the favourite tool for communication by party, the profile of the candidates with the maximum use of social media, the chronological evolution of practices over the course of the campaigns etc.

In the future, it could prove interesting to offer those tools beyond the academic sphere to political candidates themselves in order to further the understanding of practices to the benefit of both. Such a collaboration could prove beneficial as most devices for social media analytics currently available are extensions of models developed for the business sector which do not necessarily address the specific issues of the political sphere.

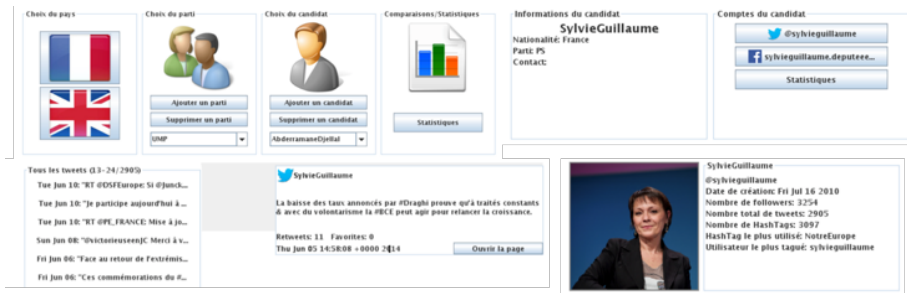

*Figure 3. Profiling a candidate's campaign on social networks*

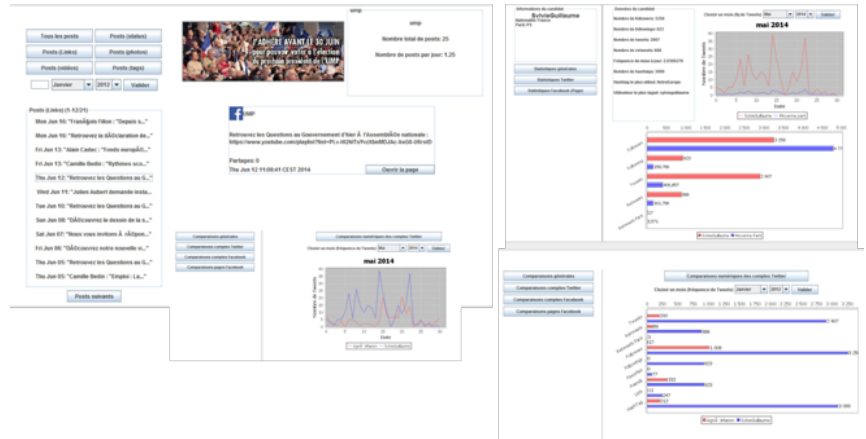

*Figure 4. Statistical profile of politician's campaigns on social networks and Internet*

*5.4 Some suggestions regarding black box, methodological and legal concerns*

Regarding the epistemological challenges raised in section 4, guidelines for addressing them are still in the making as the community involved in such projects shares feedback and experience, on platforms like Hypothèses for instance (https://fr.hypotheses.org/). Here are some contributions to this ongoing reflection.

In terms of black box effects, what the experience of the POLIWEB team suggests is the need for researchers in the humanities and social sciences to consider the investment/outcome ratio for training and to find an equilibrium between the complete delegation of technical considerations to more qualified specialists and the desire one might have of mastering all the technical aspects of a given project, i.e understanding enough of what is done by collaborators to ensure the validity of the results over the complete research process. To use a basic metaphor, when using a spreadsheet, most beginners do not necessarily know how the software goes from typing in a formula to providing a result, but they are aware of which formula will return the expected result and should be in a position to spot potential mistakes.

In terms of methodology and concerns for the reliability of results, the team found that triangulation (Denzin, 1978; Webb, Campbell, Schwartz, & Sechrest, 1966) was of capital importance. In this project, this meant the decision to combine various sources and forms of data, i.e data derived from web crawling to field data. Thus, to use Mintzberg's (Mintzberg, 1979) easily understandable if not consensual terminology, the specialist of British civilisation and politics in the project provided the 'soft' data crucial for the qualitative aspect of this triangulation through the use of interviews with candidates and campaigning teams, questionnaires, observation at party headquarters etc... and guidelines for gathering 'hard data' (i.e. those stemming from SNTs), which in turn led to statistical output.

Moreover, the participants to POLIWEB were aware of the shortcomings of mixed methods' approaches combining quantitative and qualitative data and of the debates surrounding triangulation or convergent validation. Among the major one, the risk that multiplying sources, types of data, tools and methods to ensure the validity of the findings might actually produce the opposite result if the pitfalls of each were combined. However, this decision was made partly for reasons pertaining to the research questions to be answered, i.e. not only how much was being said online, but also what and why, but also to the reluctance to be essentially dependent on the 'hard' data and the software for conclusions. Besides, it was felt that it offered more coherence with the cyclical approach referred to earlier with leads emerging from manipulation of data then tested during the interviews and in questionnaires, which in turn generated new hypotheses which could be confronted to the data.

Regarding legal issues, our work aimed at guiding the gathering, storing and scientific exploitation of online data both in France and in the UK, according to confidentiality and respect for private life. We referred to rules raised in France by the CNIL (National Commission on Freedom and Informatics) and constraints pertaining to data ownership rights born out of the emergence of big data generated online. We focused on Service Level Agreement (SLA) guided integration of heterogeneous sources with special attention paid to legal, provenance-based and privacy rules enforcement (Bennani, Vargas-Solar, Ghedira, Souza-Neto, & Carvalho, 2015), thus adapting the collection, cleaning and curation according to SLA contracts considering juridical aspects.

If a specific and comprehensive legal framework integrating all legal and ethical aspects generated by such projects is still lacking, the team also looked for guidance from different sources and in particular, the legal information platform operated by CLARIN, the European Research Infrastructure for Language Resources and Technology (https://www.clarin.eu/content/legal-information-platform) as well as the *Questions Etiques et Droit en SHS* blog page (https://ethiquedroit.hypotheses.org/) provided by a working group of professionals (jurists and librarians) from Aix-Marseille university.

**6. Conclusions and Future Work**

The increasing use of ICTs by political candidates generates an unprecedented volume of technologically-mediated online information commonly referred to as "Big Data" which presents a number of challenges. The necessity to manage and organise such large quantities of digital contents into meaningful items is among them and has become increasingly pressing for social scientists desirous to make use of such wealth of data but facing difficulties in matters related to gathering, storing and retrieving information on so large a scale.
We aimed to map out the diversity of uses of the internet for political campaigning in a variety of contexts and thus provide a comprehensive overview of practices and expectations from candidates. Identifying potential applications and partners for this type of development shall be another objective, as well as developing tools to facilitate political science analysis in the context of future campaigns.

We adopted a classic workflow in Data Science which is less classic in Humanities, and includes activities addressing data collection, cleaning and curation. We used the notion of view to curate data collections, that is to maintain them and make them easy to explore and usable for digital Humanities. These activities of the workflow are guided by SLA criteria particularly juridical ones which control the way these processing phases are performed according to the type of data, the conditions in which they are produced and consumed.

The workflow activities consume computing, memory and storage resources at different scales depending on the volume of data and the complexity of the algorithms used. So, some of them are implemented under the map-reduce programming model and implemented and executed on cluster like architectures, using existing tools. Our first contribution in this paper regards the strategies used for characterizing and inferring data content through the notion of view. Some inference had to deal with uncertainty which we addressed associating accuracy probabilities to inferences so as to guide the data scientist in her further data analytics design.

To conclude, as described in this paper, the project POLIWEB generated a number of challenges for which solutions could not always be found. However, the process so far has been very instructive and the mistakes we made most of all. However, mistakes can be costly in many ways and the team welcomes this opportunity to share its experiences so as to enable colleagues faced with similar difficulties to make as few of them as possible in the future, in the belief that sharing doubts and reflections from case studies can serve a purpose for the scientific community.